\documentclass[12pt]{article}

\usepackage[a4paper,left=30mm,right=20mm,top=20mm,bottom=20mm]{geometry}
\usepackage{graphics}
\usepackage{amsmath}
\usepackage{epsfig}
\usepackage{cite}
\usepackage{xcolor}

\newcommand{\bea}{\begin{eqnarray}}
\newcommand{\eea}{\end{eqnarray}}
\newcommand{\be}{\begin{equation}}
\newcommand{\ee}{\end{equation}}

\newcommand{\Li}{\mathop{\mathrm{Li}}\nolimits}

\title{On $x$-independence of $R^Q = F_L^Q/F_2^Q$ ratio at low $x$}

\author{N.A.~Abdulov$^{1}$, A.V.~Kotikov$^{1}$, A.V.~Lipatov$^{1,2}$}

\begin{document}

\maketitle

\begin{center}
 
{\it $^{1}$Joint Institute for Nuclear Research, 141980, Dubna, Moscow region, Russia}\\
{\it $^{2}$Skobeltsyn Institute of Nuclear Physics, Lomonosov Moscow State University, 119991, Moscow, Russia}

\end{center}

\vspace{0.5cm}

\begin{center}

{\bf Abstract }

\end{center}

\indent
We give predictions for the ratio $R^Q(x,Q^2)=F_L^Q(x,Q^2)/F_2^Q(x,Q^2)$ at small values of Bjorken
variable $x$ in the first three orders of perturbation theory.
We show an approximate $x$-independence of $R^Q(x,Q^2)$ at low $x$ and 
non-large $Q^2$ values ($Q^2 \leq 8\div 10 m^2_{Q}$), 
  irrespectively 
on the gluon density in a proton used in the calculations.
This observation could be useful in subsequent phenomenological studies of the 
heavy flavor production at future lepton-hadron and hadron-hadron colliders.

\vspace{1.0cm}


\newpage

\section{Introduction} \indent

An important new data on the cross sections for the
open charm and beauty production in deep inelastic electron-proton scattering (DIS) have appeared \cite{H1+ZEUS:2018} by combining the results of research from
the H1 and ZEUS collaborations at HERA.
Measurements have shown that the production of heavy flavor $Q$ in DIS occurs predominantly 
due to the
photon-gluon fusion process $\gamma g \to Q\overline{Q}$
and therefore depend strongly on gluon density in the proton and 
mass $m_Q$ of produced heavy quark\footnote{In $\overline{MS}$-scheme $m_Q$ is $Q^2$-dependent. Since the $Q^2$-dependence is rather slow, we will use below $m_Q=m_Q(Q^2=m_Q^2)$.}.
Theoretical studies usually serve to confirm that available HERA data can be 
described by perturbative heavy flavor generation in QCD (see, for example, review\cite{Gao:2017yyd} and 
references therein).
Further investigations are planned at future lepton-hadron and hadron-hadron 
colliders\cite{Amoroso:2022eow}, such as LHeC, FCC-eh and FCC-hh (see also\cite{AbelleiraFernandez:2012cc} for review),
where the measurements can be performed with much increased precision and extended 
to much smaller $x$ and high $Q^2$ values.

In our previous consideration\cite{KLSZ,Kotikov:2021yzb} we have analyzed
latest experimental data\cite{H1+ZEUS:2018}
%on the charm and beauty production at HERA
taken by the H1 and ZEUS Collaborations at HERA.
In particular, we have studied %\footnote{We have used so-called $k_T$-factorization\cite{?} (or high-energy factorization\cite{?}) approach
%valid at low $x$ (see review\cite{?} for more information).}
the heavy quark contributions to the 
proton structure function (SF) $F_2(x, Q^2)$
and reduced charm and beauty cross sections $\sigma^{c\overline{c}}_{\rm red}(x, Q^2)$ and 
$\sigma^{b\overline{b}}_{\rm red}(x, Q^2)$
measured mostly at small values of the Bjorken variable $x$ in a wide region of $Q^2$\cite{69}.
An additional important result of the evaluations \cite{Kotikov:2021yzb}
%calculations\cite{?}
is that the compact analytical expressions for these DIS coefficient 
functions have been presented 
%in the conventional on-shell limit
up to next-to-next-to-leading order (NNLO) accuracy. 
These expressions were used very recently to investigate the top quark production in 
the FCC-he kinematical regime\cite{Boroun:2021skz}.
Here we continue our study\cite{KLSZ,Kotikov:2021yzb}.
Using the previously derived expressions for DIS
coefficient functions, we 
study the ratio $R^{Q}(x,Q^2) = F^Q_L(x,Q^2)/F^Q_2(x,Q^2)$ 
in the first three orders of perturbation theory
We demonstrate an approximate $x$-independence
%  very slow $x$-dependence 
  of this ratio for non-large $Q^2$ values, namely, $Q^2 \leq 8\div 10 m_Q^2$. Moreover, we show 
  a very slow ratio's dependence on the choice of used gluon density.
  %}

\section{Basic formulas} \indent

It is well known that the differential cross section $\sigma^{Q \bar Q}$
of heavy flavor production in DIS
can be presented in the simple form:
\be
   {d^2 \sigma^{Q \bar Q} \over dx dy} = {2 \pi \alpha^2 \over x Q^4} \left[ \left( 1 - y + {y^2\over 2}\right) F_2^{Q}(x,Q^2) - {y^2 \over 2} F_L^{Q}(x, Q^2)\right],
  \label{sigmaff}
%\eea
\ee
\noindent
where
$x$ and $y$ are the usual Bjorken scaling variables.

Suggesting that the SFs $F_k^{Q}(x, Q^2)$ ($k=2,L$)  are driven at small $x$ by gluons (see \cite{Gao:2017yyd}),
%all heavy quarks are dynamical
it is easily to obtain the following
%LO
results
(see, for example,\cite{Illarionov:2008be,IllaKo}):
%(with the same approximation, as in the previous subsection):
\begin{equation}
  F_k^Q(x,Q^2) = C_{k,g}(x,Q^2,m^2_Q) \, \otimes f_g(x,Q^2),
\label{eq:pm3}
\end{equation}
where the symbol
%SFs, obtained in the generalized DAS approach, were marked as $\hat{F}_k^Q(x,Q^2)$. Here
$\otimes $ marks
%is
the Mellin convolution:
\be
C_{k,g}(x,Q^2,a_Q) \, \otimes f_g(x,Q^2) \equiv
\int_{x}^{b} \, \frac{dy}{y} \, C_{k,g}(y,Q^2,a_Q) \, f_g(x/y,Q^2),
\label{otimes}
\ee
\noindent
with
%$\Theta(x_0-x)$ is the Heaviside step function and 
 \be
b~=~\frac{1}{1+4a_Q},~~~a_Q=\frac{m^2_Q}{Q^2}.  \,
%\nonumber
 \label{x1}
 \ee

\noindent
So, the ratio $R^Q(x,Q^2)$ can be presented as
\begin{equation}
  R^Q(x,Q^2) = \frac{F_L^Q(x,Q^2)}{F_2^Q(x,Q^2)} =
  %\approx\
%  \frac{M_{L,g}(1,Q^2,m_f^2)}{M_{2,g}(1,Q^2,m_f^2)},
\frac{C_{L,g}(x,Q^2,a_Q) \, \otimes f_g(x,Q^2)}{C_{2,g}(x,Q^2,a_Q) \, \otimes f_g(x,Q^2)} \, ,
  \label{eq:rio}
\end{equation}
%which is very useful for practical applications (see the section 4.3).
where $f_g(x,Q^2)$ is the gluon density in a proton\footnote{We use the gluon density multiplied by $x$.}.
%given in (\ref{8.02}).
In fact, the ratio $R^Q(x,Q^2)$ depends slowly on
 non-perturbative input $f_g(x,Q^2)$, which contribute to the both numerator and denominator 
 of the ratio $R^Q(x,Q^2)$.

   As we already noted in Introduction,
we study the ratio  $R^Q(x,Q^2)$ at small values of the Bjorken variable $x$.
 Using the 
  results \cite{Catani:1996sc,Catani:1992zc,Kawamura:2012cr}, below we give formulas for the high energy asymptotics of collinear coefficient functions of the
  heavy quark production process in the first three orders of the perturbation theory.
  
 \subsection{Wilson coefficients} \indent

 Taking the Wilson coefficient in the $m$ order of perturbation theory, we have
 \be
 C_{k,g}(x) ~=~ e_Q^2  a_s(\mu^2) \,  \overline{C}^{(m)}_{k,g}(x),~~  \overline{C}^{(m)}_{k,g}(x)=  B^{(0)}_{k,g}(x,a)+
 \sum_{l=1}^m a_s^l \, B^{(l)}_{k,g}(x,a,\mu),  (m=0,1,2)\, ,
   \label{6a}
   \ee
   where $e_Q$ is the charge of heavy quark $Q$ and $a_s(\mu^2)=\alpha_s(\mu^2)/(4\pi)$ is a strong coupling constant. Moreover, $\mu$ is the renormalization scale, which is usully
   taken in the following form \cite{Gao:2017yyd,AbelleiraFernandez:2012cc}
\be
\mu^2 = \mu^2_Q \equiv Q^2+4m_Q^2\,.
\label{muQ}
\ee
   
%   \subsubsection*{LO} \indent

 {\bf 1.}  The LO Wilson coefficients have the following form:
%   where
 \begin{eqnarray}
&&B^{(0)}_{2,g}(x,a) =  -2x\beta \,
\Biggl[ \biggl(1-4x(2-a)(1-x) \biggr)
\nonumber \\
&&\hspace{2cm} - 
\biggl(1-2x(1-2a)+2x^2(1-6a-4a^2) \biggr) \;
L(\beta) 
%\ln\frac{1+\beta }{1-\beta }  
\Biggr]\,,
 \nonumber \\
%\label{6.6}\\
&&B^{(0)}_{L,g}(x,a) = 8x^2\beta \,
\Biggl[ (1-x)
%\nonumber \\
  - 2xa \;
  L(\beta)
%\ln\frac{1+\beta }{1-\beta }  
\Biggr] \, ,
 \label{6.7}
 \end{eqnarray}
where
\be
\beta^2=
%\overline{\beta}^2(b=0)=
1-\frac{4ax}{(1-x)},~~
L(\beta) = \frac{1}{\beta} \,
\ln\frac{1+\beta }{1-\beta } \, .
\label{Lbe}
\ee

So, for the  ratio $R^Q_{\rm LO}(x,Q^2)$ we have:
\begin{equation}
R^Q_{\rm LO}(x,Q^2) =
\frac{B^{(0)}_{L,g}(x,a_Q) \, \otimes f_g(x,Q^2)}{B^{(0)}_{2,g}(x,a_Q) \, \otimes f_g(x,Q^2)} \, .
\label{eq:ri}
\end{equation}

{\bf 2.}  
The NLO coefficient functions $B^{(1)}_{k,g}(x,a,\mu)$ of photon-gluon fusion subprocess are rather lengthy and not published in
print; they are only available as computer codes\cite{Laenen:1992zk}.
Following \cite{Illarionov:2008be,IllaKo}, 
%For the purpose of this letter,
it is sufficient to work in the high energy
regime, defined by $x\ll1$, where they assume the compact form\footnote{Following Ref. \cite{Catani:1996sc}, we will use the case $M^2=4m^2$ in the colliner approach. We would like to note that in the
  original papers\cite{Catani:1992zc,Kawamura:2012cr} the scale $M^2=m^2$ has been used, which is inconsistent with the results in eqs.
  %(\ref{eq:ja}),
  (\ref{eq:nlo1}) and (\ref{eq:nnlo1}).}\cite{Catani:1996sc,Catani:1992zc,Kawamura:2012cr}:
  \begin{equation}
  B_{k,g}^{(1)}(x,a,\mu)=\beta \bigl[R_{k,g}^{(1)}(1,a) + 4 C_A B_{k,g}^{(0)}(1,a) L_{\mu} \bigr] \, ,~~ L_{\mu}=\ln\frac{M^2}{\mu^2} \, , M^2=4m^2, \,
\label{eq:nlo}
\end{equation}
with
\bea
&&R_{2,g}^{(1)}(1,a)~=~\frac{8}{9}C_A[5+(13-10a)J(a)+6(1-a)I(a)],~~
%R_{k,g}^{(2)}(1,a)~=~-4 C_A M_{k,g}^{(0)}(1,a),
\nonumber \\
&&R_{L,g}^{(1)}(1,a)~=~-\frac{16}{9}C_A b \,\bigl[1-12a-(3+4a(1-6a))J(a)+12a(1+3a)I(a)\bigr]
\label{eq:nloA}
\eea
and
  %\bea
\be
B_{2,g}^{(0)}(1,a)=
%&=&
\frac{2}{3} \, [1+2(1-a)J(a)],~~
%\nonumber\\
B_{L,g}^{(0)}(1,a)=
%&=&
\frac{4}{3} \, b \, [1+6a-4a(1+3a)J(a)],
\label{Bk0}
\ee
%\end{eqnarray}
where\footnote{The functions $J(a)$ and $I(a)$ in
  %(\ref{eq:ja}) and
  (\ref{eq:nlo1}) coincide with ones in 
  \cite{Illarionov:2008be} and differ from ones in \cite{Catani:1992zc,Kawamura:2012cr,Catani:1996sc} by an additional factor
  $4a$. The function $K(a)$ in (\ref{eq:nnlo1}) coincides with the combination $4a[K(a)+\ln(4ab)I(a)]$.}
\be
J(a) = - \sqrt{b}\ln t,~~
I(a)=-\sqrt{b}\left[\zeta(2)+\frac{1}{2}\ln^2t-\ln(ab)\ln t+2\Li_2(-t)\right],~~ t=\frac{1-\sqrt{b}}{1+\sqrt{b}},
\label{eq:nlo1}
\ee
%\eea
with
\be
\Li_2(x)=-\int_0^1\,\frac{dy}{ y}\, \ln(1-xy)
\label{Li2}
\end{equation}
being the dilogarithmic function.
We would like to note that $B_{k,g}^{(0)}(1,a)$ are the first moments 
of the LO Wilson coefficients $B_{k,g}^{(0)}(x,a)$ (see
%~(\ref{6.6}) and~
(\ref{6.7})):
\begin{equation}
B_{k,g}^{(0)}(n,a)=\int_0^{b}dx\,x^{n-2} B_{k,g}^{(0)}(x,a) \, .
\label{eq:melB}
\end{equation}
\noindent
So, at NLO we have:
\begin{equation}
R^Q_{\rm NLO}(x,Q^2) =
\frac{\overline{C}^{(1)}_{L,g}(x,a_Q) \, \otimes f_g(x,Q^2)}{\overline{C}^{(1)}_{2,g}(x,a_Q) \, \otimes f_g(x,Q^2)}\,.
  \label{eq:tri}
\end{equation}

{\bf 3.}  
In the high energy regime the coefficient $B^{(2)}_{k,g}(x,a,\mu)$ has 
   %  , defined by $x\ll1$, where they assume
   the compact form:
%cite{Catani:1992zc}
\begin{equation}
  B_{k,g}^{(2)}(x,a,\mu)=\beta \, \ln(1/x) \,\bigl[R_{k,g}^{(2)}(1,a) + 4 C_A R_{k,g}^{(1)}(1,a) L_{\mu} + 8 C_A^2 B_{k,g}^{(0)}(1,a) L^2_{\mu} \bigr] + O(x^0)\, ,
\label{eq:nnlo}
\end{equation}
with
\bea
%\bea&&
&&R_{2,g}^{(2)}(1,a)~=~\frac{32}{27}C_A^2 \, [46+(71-92a)J(a)+3(13-10a)I(a)-9(1-a)K(a)],~~\nonumber \\
&&R_{L,g}^{(1)}(1,a)~=~\frac{64}{27}C_A^2 b\,\bigl\{34+240a-[3+136a+480a^2]J(a)
+3[3+4a(1-6a)]I(a) \nonumber \\
&& \hspace{2cm} +18a(1+3a)K(a)\bigr\} \, ,
\label{eq:nnloA} 
\eea
where
(\ref{eq:nlo1}), respectively, and
\bea
  K(a)&=&-\sqrt{b}\biggl[ 4\Bigl(\zeta(3)+\Li_3(-t)-\Li_2(-t)\ln t-2S_{1,2}(-t) \Bigr)+ 2\ln(ab)
    \nonumber \\
    &&\times \Bigl(\zeta(2)+2\Li_2(-t)\Bigr) - \frac{1}{3}\ln^3t-\ln^2(ab)\ln t+\ln(ab)\ln^2 t\biggr],
\label{eq:nnlo1}
\eea
where $t$ is given in (\ref{eq:nlo1})
%(\ref{eq:ja})
and
%$\zeta(2)=\pi^2/6$ and
\be
\Li_3(x)=\int_0^1\,\frac{dy}{ y}\, \ln(y) \, \ln(1-xy),~~S_{1,2}(x)= \frac{1}{2} \, \int_0^1\,\frac{dy}{ y}\, \ln^2(1-xy),~
\label{Li3}
\end{equation}
 are the trilogarithmic function $\Li_3(x)$ and Nilsen Polylogarithm $S_{1,2}(x)$ (see\cite{Devoto:1983tc}).
The results for $K(a)$ in the form of harmonic Polylogarithms\cite{Remiddi:1999ew} can be found
in\cite{Kawamura:2012cr}.
 
So, at NNLO we have:
\begin{equation}
R^Q_{\rm NNLO}(x,Q^2) =
\frac{\overline{C}^{(2)}_{L,g}(x,a_Q) \, \otimes f_g(x,Q^2)}{\overline{C}^{(2)}_{2,g}(x,a_Q) \, \otimes f_g(x,Q^2)} \,.
  \label{eq:tri}
\end{equation}

\subsection{Gluon density in a proton} \indent

In the previous paper \cite{Kotikov:2021yzb}, the gluon density $f_g(x,Q^2)$ coming from the generalized double asymptotic scaling (gDAS) approach
\cite{Q2evo,Cvetic1} has been applied.
The usage of the gluon density leads to a good description of experimental data for the 
SFs $F_2(x,Q^2)$ and $F_2^c(x,Q^2)$ (see, for example, \cite{Cvetic1,Kotikov:2016oqm} and \cite{Illarionov:2008be,IllaKo}, respectively). Moreover,
the parton densities of the generalized DAS approach have been used successfully \cite{KLSZ,Kotikov:2021yzb,Abdulov:2022itv}
to generate the TMD parton densities (see the recent paper \cite{Abdulov:2021ivr}), applied in the $k_T$-factorization approach
\cite{2,3}, which is very useful in partially to study $F_2^Q(x,Q^2)$ 
%$F_2^c(x,Q^2)$ and $F_2^b(x,Q^2)$
and the corresponding (reduced)
%both
charm and beauty cross sections $\sigma^{c\overline{c}}_{\rm red}$ and $\sigma^{b\overline{b}}_{\rm red}$.

Here, with a purpose to implement more information from experimental data, 
%\textcolor{blue}{
  additionally
%}
we will use the gluon density, corresponding to H1 parametrization \cite{H1:2001ert}
(see also \cite{Surrow:2001ghf,Kotikov:2002fd})
for the SF
%structure function
$F_2(x,Q^2)$:
\be
F_2^{\rm H}(x,Q^2)=C_{\rm H}\, x^{-\lambda_{\rm H}(Q^2)},~~\lambda_{\rm H}(Q^2)= a_{\rm H}\,\ln\left(\frac{Q^2}{\Lambda^2_{\rm H}}\right)\,,
\label{F2H1}
\ee
with the following parameters \cite{H1:2001ert}
\be
C_{\rm H}=0.18,~~a_{\rm H}=0.048,~~\Lambda_{\rm H}=292 \, \mbox{MeV}^2\,.
\label{F2H1pa}
\ee

Assuming the basic contribution of $f_g(x,Q^2)$ to the SF
%structure function
$F_2(x,Q^2)$ at small $x$ values, we can suggest it in the following form
\be
f_g^{\rm H}(x,Q^2)=C^{g}_{\rm H}\, x^{-\lambda_{\rm H}(Q^2)},
\label{F2H1}
\ee
with some unknown $C^{g}_{\rm H}$, which value is not so important for the present investigations.

\section{Resume} \indent

\begin{figure}
\begin{center}
\includegraphics[width=7cm]{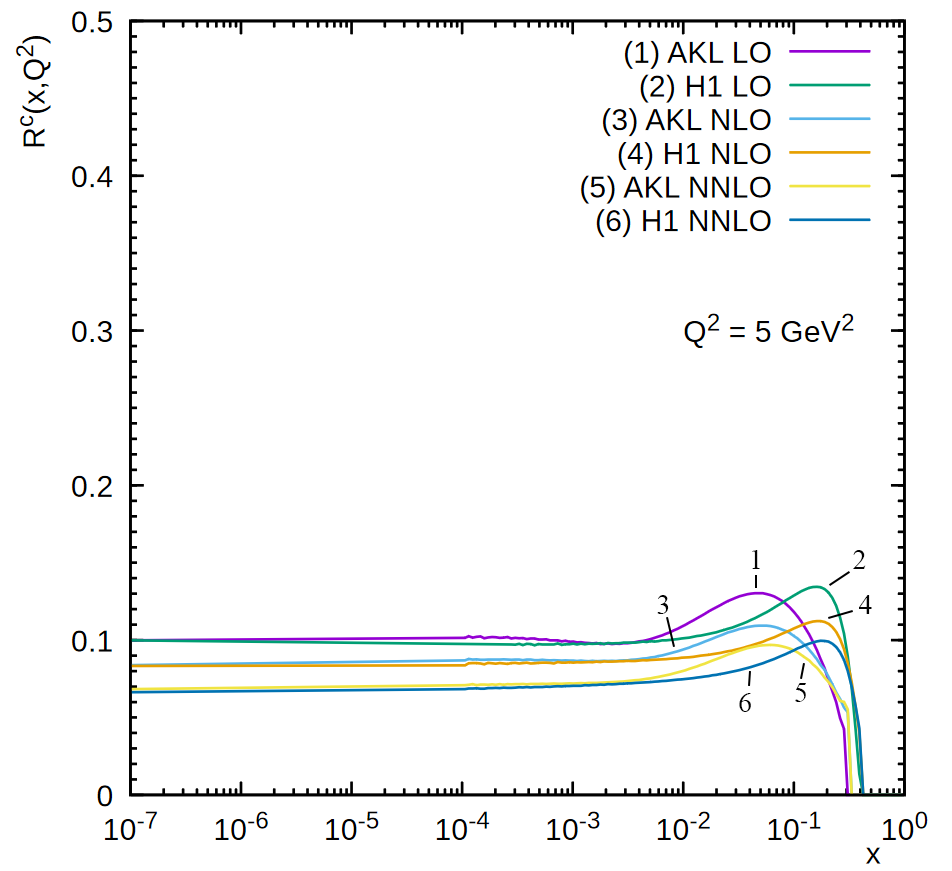}
\includegraphics[width=7cm]{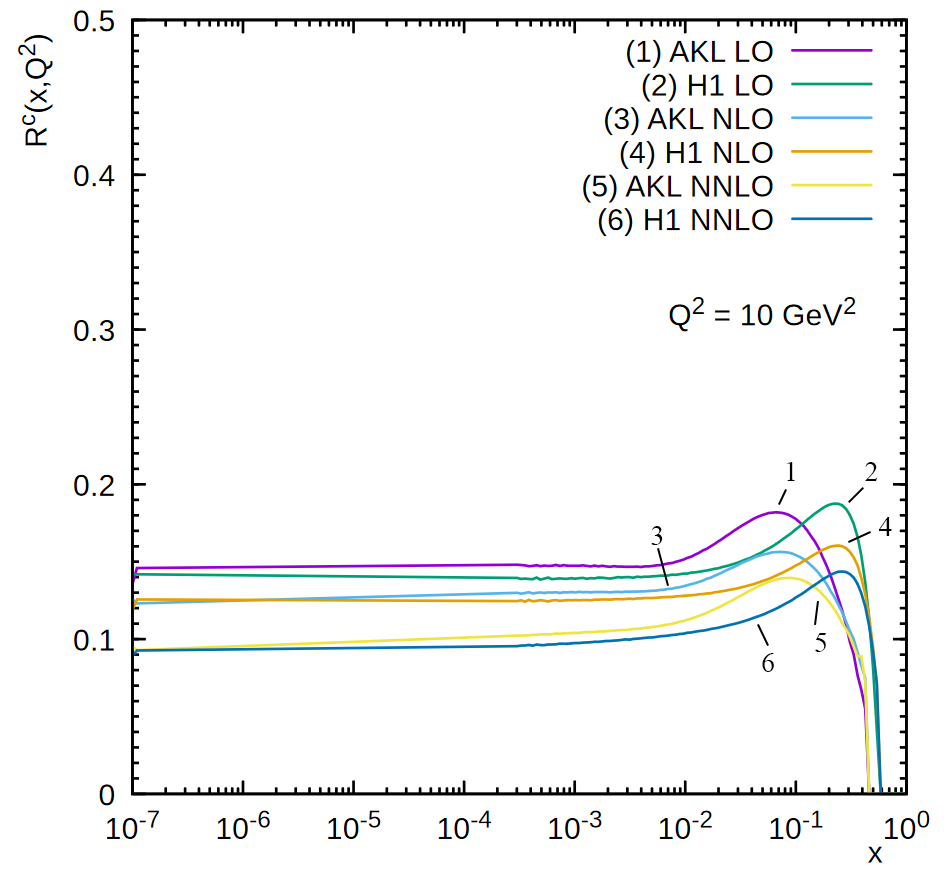}
\includegraphics[width=7cm]{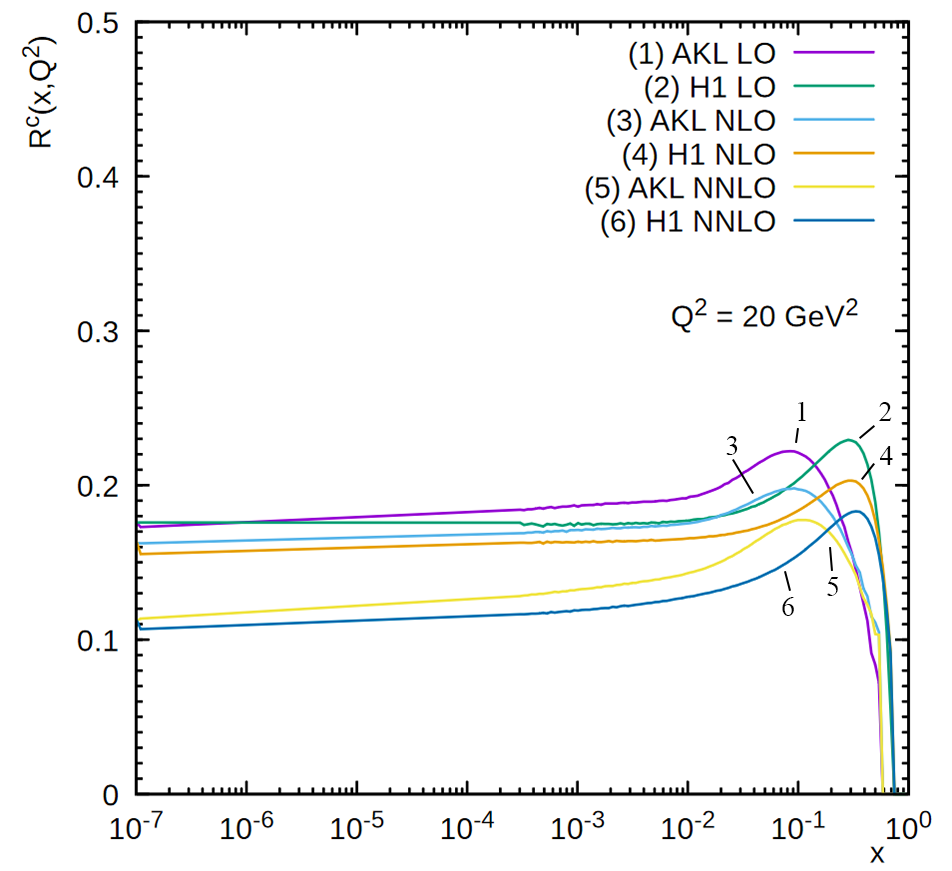}
\includegraphics[width=7cm]{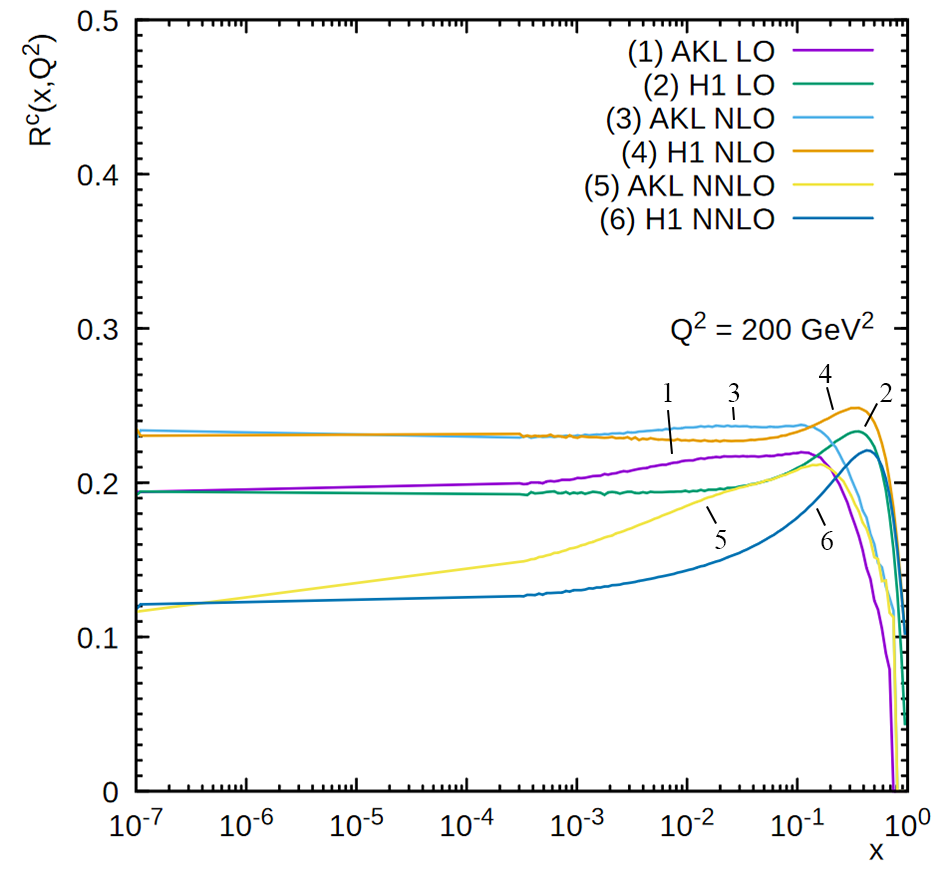}
\caption{The ratio $R^c(x,Q^2)$ as a function of $x$ at different $Q^2$ values. The purple (1), azure (3) and yellow (5) curves correspond 
  to the results obtained with AKL gluon density function \cite{Abdulov:2021ivr} at LO, NLO and NNLO, respectively.
  The green (2), orange (4) and blue (6) curves correspond 
to the results obtained with H1 parametrization from Eq. (\ref{F2H1}) at LO, NLO and NNLO, respectively.}
\label{figRc}
\end{center}
\end{figure}

\begin{figure}
\begin{center}
\includegraphics[width=7cm]{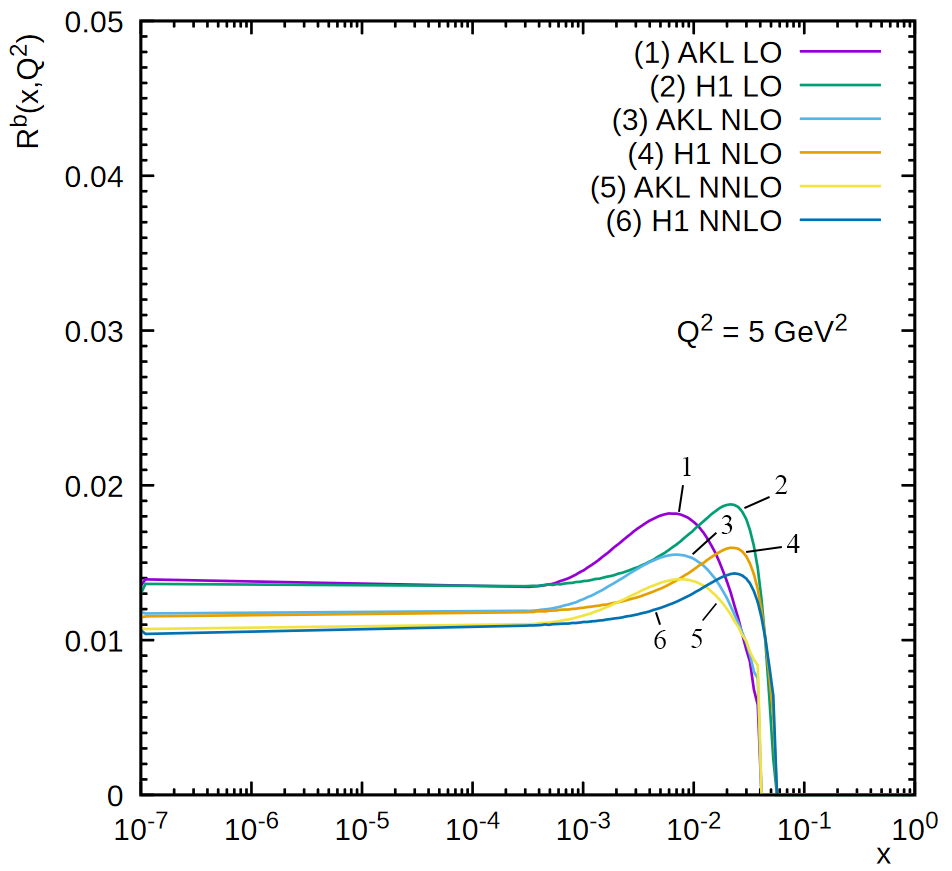}
\includegraphics[width=7cm]{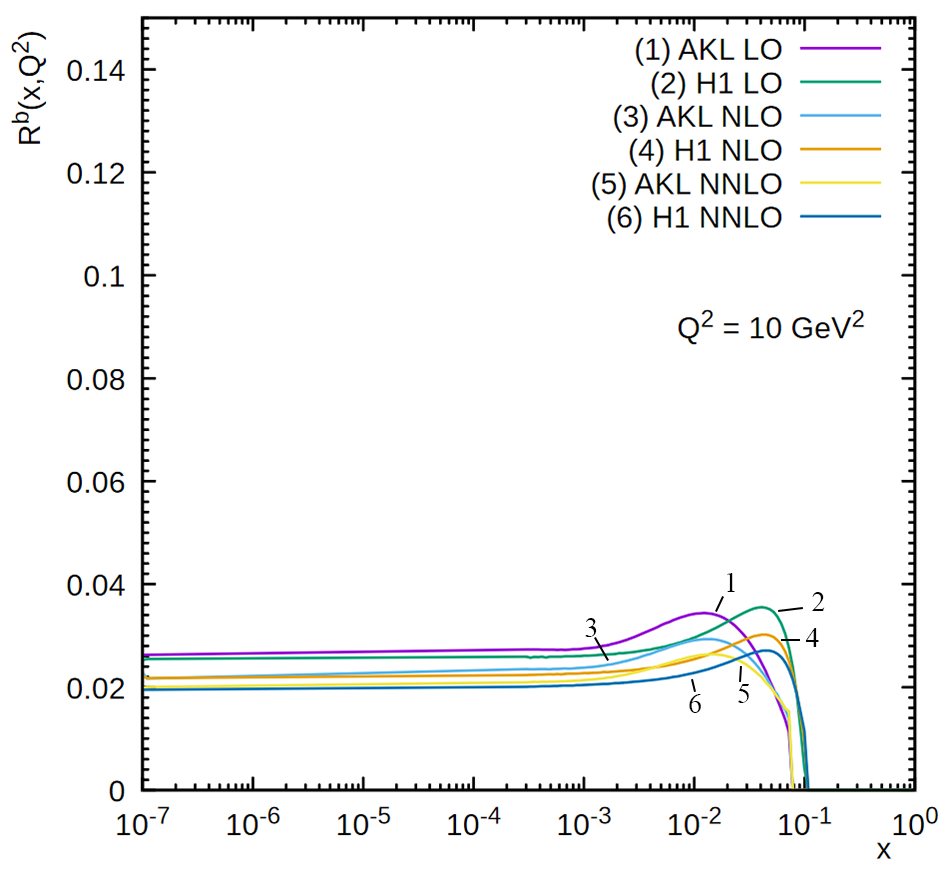}
\includegraphics[width=7cm]{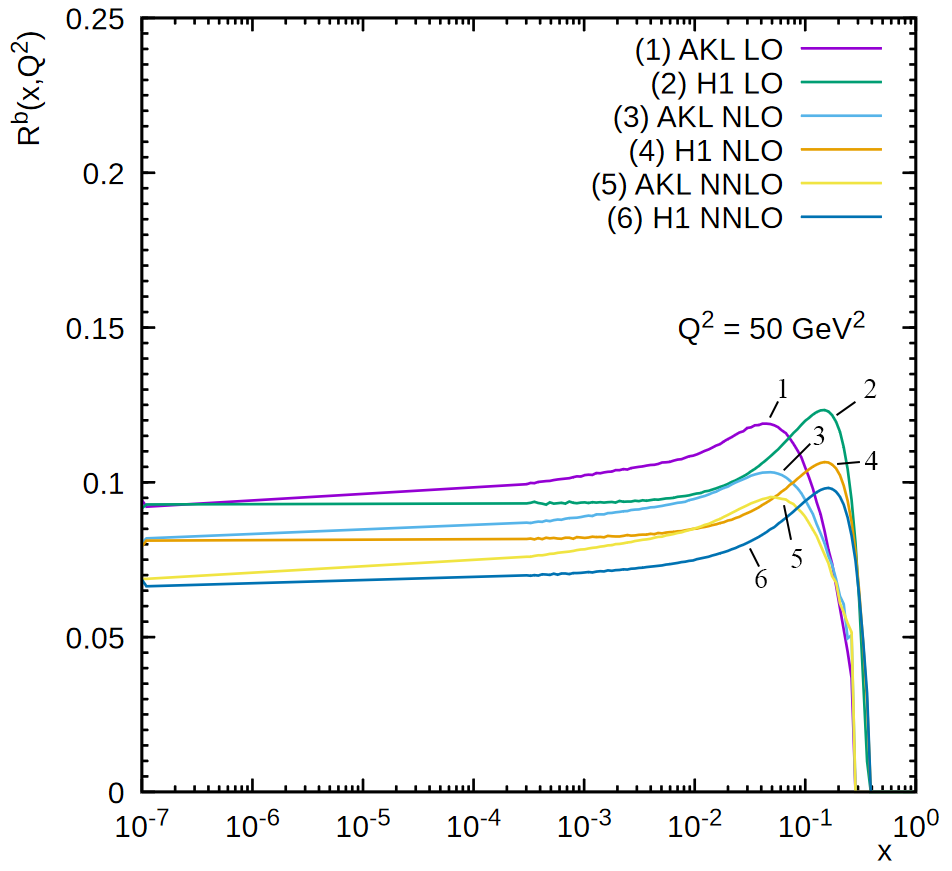}
\includegraphics[width=7cm]{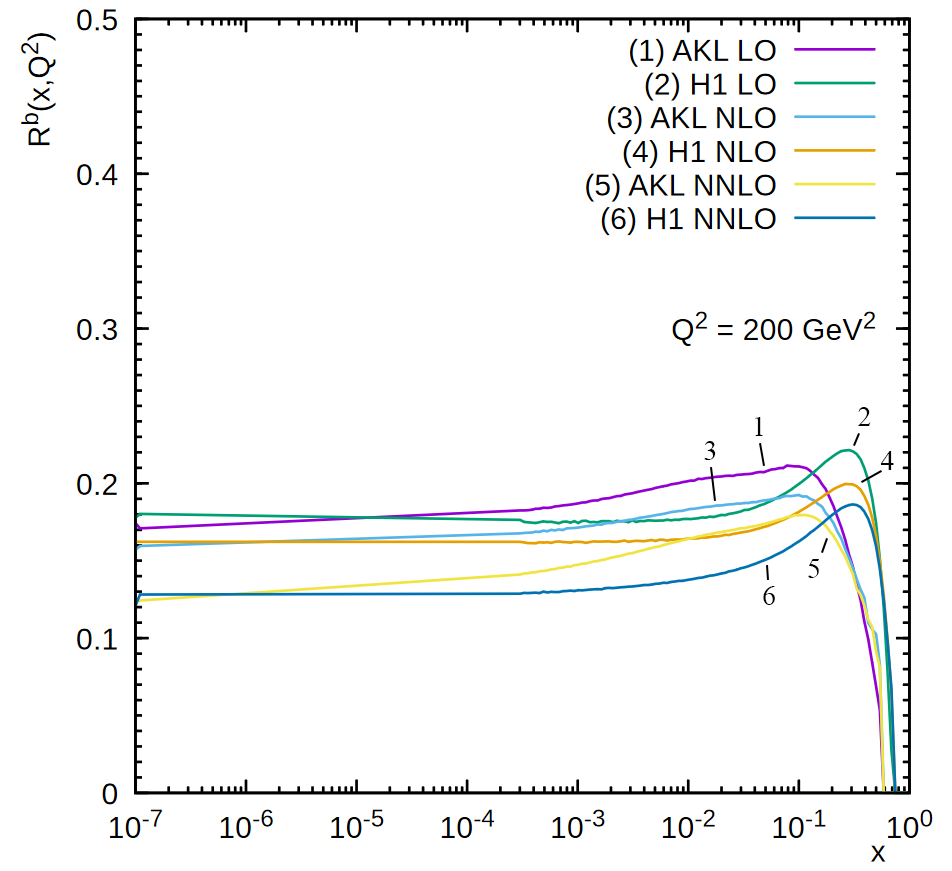}
\caption{The ratio $R^b(x,Q^2)$ as a function of $x$ at different $Q^2$ values. Notation of curves is the same as in Fig.~\ref{figRc}.}
\label{figRb}
\end{center}
\end{figure}

\begin{figure}
\begin{center}
\includegraphics[width=7cm]{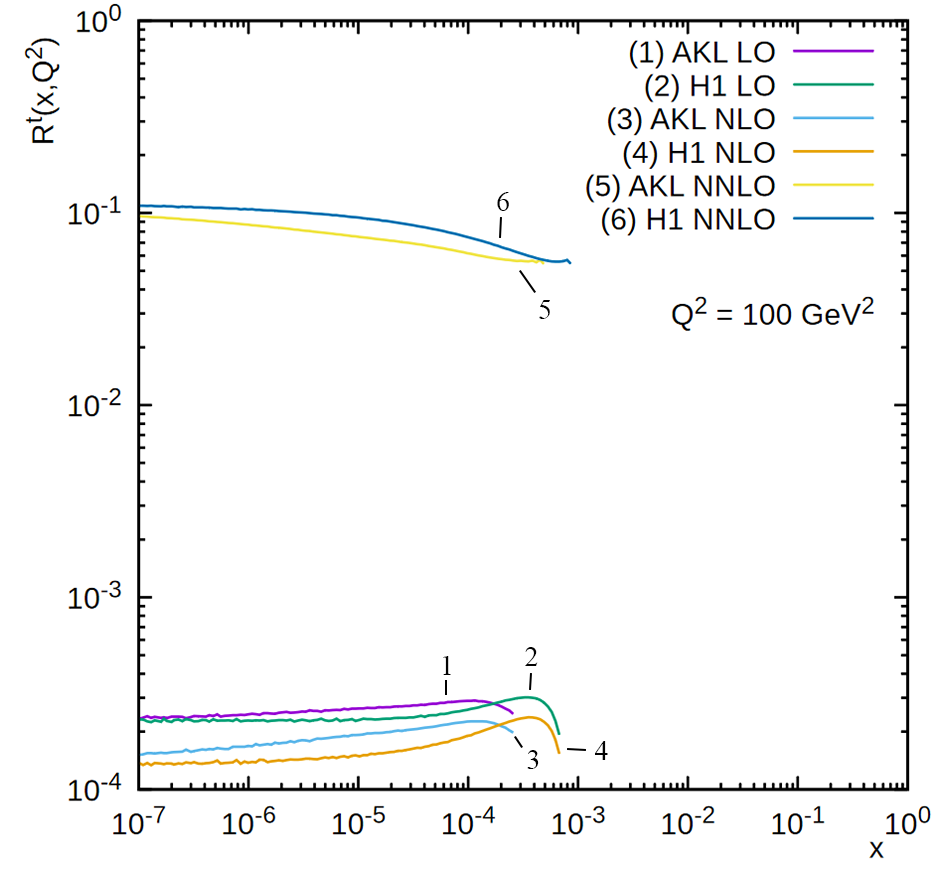}
\includegraphics[width=7cm]{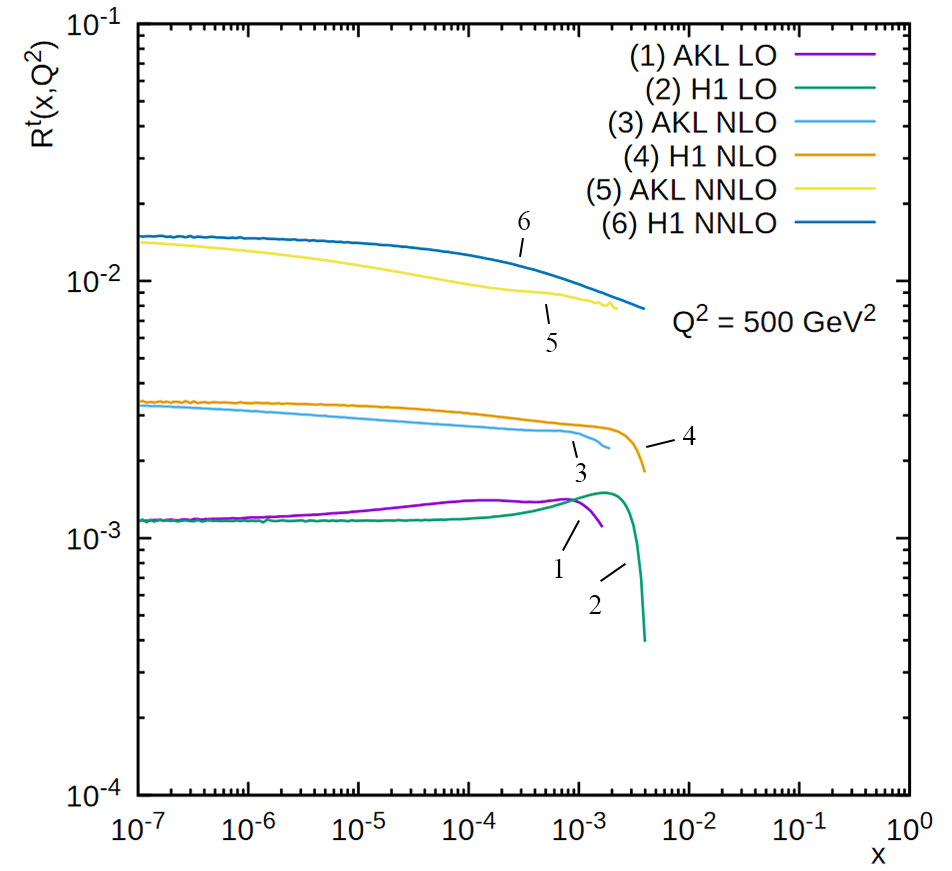}
\includegraphics[width=7cm]{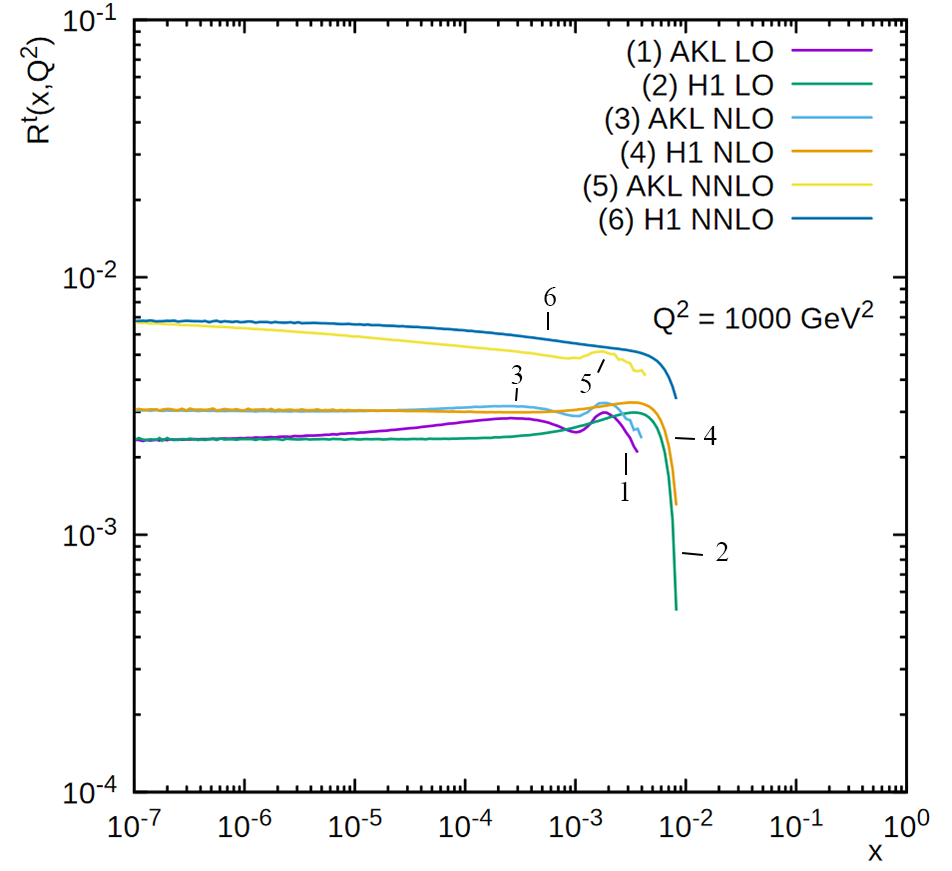}
\includegraphics[width=7cm]{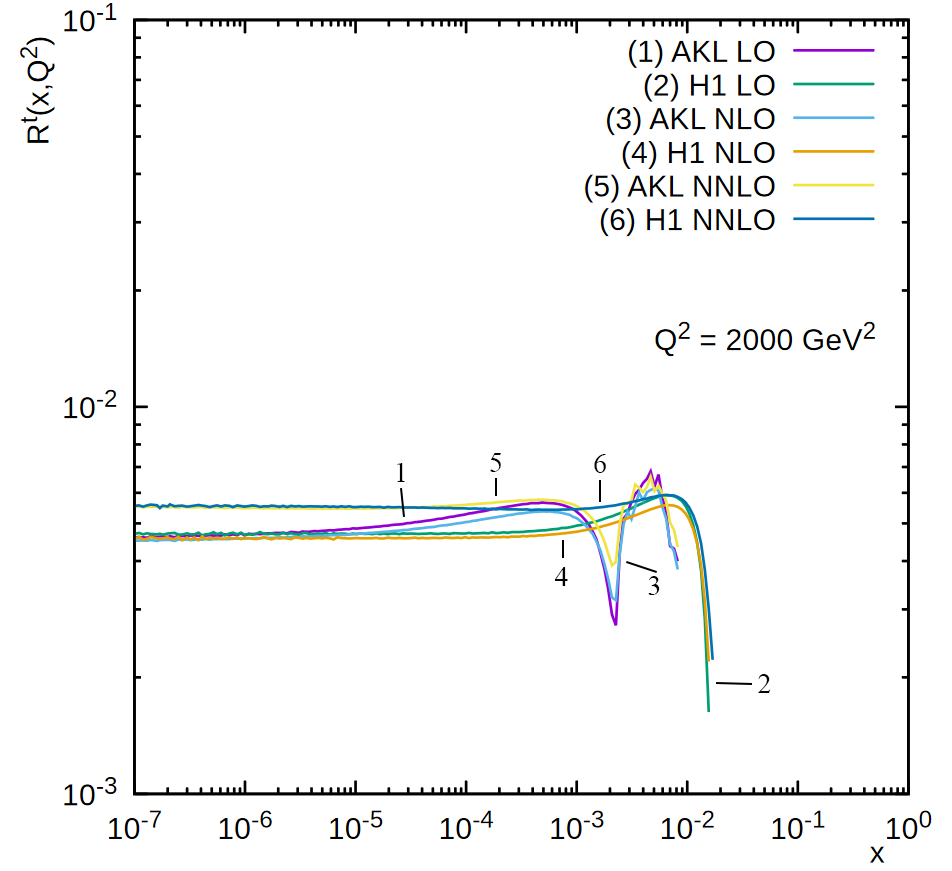}
\caption{The ratio $R^t(x,Q^2)$ as a function of $x$ at different $Q^2$ values. Notation of curves is the same as in Fig.~\ref{figRc}.}
\label{figRt}
\end{center}
\end{figure}

The predictions are shown in Figs.~\ref{figRc} --- \ref{figRt}. Our calculations show a flat ($x$-independent) behavior of $R^{Q}(x,Q^2)$ for small values of $x$ ($x \leq 10^{-2}$) and at not very
large $Q^2$-values ($Q^2 \leq 8\div 10 m^2_{Q}$).
For larger values of $Q^2$, some dependence on $x$ appears, especially in the NNLO case, which  has strongly singular
coefficient functions $\sim 1/(n-1)^2$ and
%, which
leads to the strengthening $\sim \ln x$ in the NNLO contributions at low $x$ (see Appendix B in Ref. \cite{Kotikov:2021yzb}).
%\textcolor{red}{
In the case of $c$ and $b$ quarks, such flat behavior is in agreement with one observed earlier \cite{Kotikov:2021yzb,KLZ} within the framework of the
  $k_T$-factorization approach \cite{3,2}. In the case of $t$ quark, this flat behavior was recently 
discussed in the second paper of Ref. \cite{Boroun:2021skz}.
%}
Very large NNLO contribution to $R^{t}(x,Q^2)$ for $Q^2 \ll 4m_t^2$ indicates a strong need for resummation for small $x$
    (see, for example,\cite{Catani:1996sc,Catani:1992zc,Resumm1,Resumm2} and discussions therein), but that is beyond the scope of this short paper.

%\textcolor{red}{
As one can see in Figs.~\ref{figRc} --- \ref{figRt},
the obtained results for $R^{Q}(x,Q^2)$ are very similar for both gluon densities even though they have very different low $x$ asymptotics.
So, this shows a very slow dependence of the ratio $R^{Q}(x,Q^2)$ on used gluon density.
%}
It is interesting that such flat behavior observed for $R^{Q}(x,Q^2)$ at the first three orders of perturbation theory strongly contradicts the behavior
of the ratio $R=F_L/(F_2-F_L)$, which becomes strongly $x$-dependent in higher orders of perturbation theory (see e.g., \cite{Gao:2017yyd})
An attempt to resum the large terms in the perturbation theory coefficients leads \cite{Kotikov:1993yw} to large values of
%an $x$-dependent
the argument of the strong
coupling constant.

We hope that the observed 
approximate $x$-independence of $R^{Q}(x,Q^2)$ at low $x$ will be useful in further phenomenological studies of 
the (reduced) cross sections  $\sigma^{Q\overline{Q}}_{\rm red}$ and SFs
$F^Q_2(x,Q^2)$ at future lepton-hadron and hadron-hadron colliders, such as LHeC, FCC-eh and FCC-hh.\\

{\sl Acknowledgements}.
Researches described in Section 2 were supported by the Russian Science Foundation under
grant 22-22-00387. Studies described and performed in Section 3 were supported by the
Russian Science Foundation under grant 22-22-00119.

%\end{document}

\newpage

%\begin{figure}
%\begin{center}
%\includegraphics[width=15cm]{Rc5.png}
%\caption{The charm and beauty contribution to the ratio $\hat{R}^c(x, Q^2)$
  %proton structure function $F_2(x, Q^2)$
%  as a function of $x$ calculated at
 % different $Q^2$ values. Notation of green and yellow solid curves is the same as in Fig.~1.
 % The collinear results for $\hat{R}^c_{\rm LO}(x, Q^2)$, $\hat{R}^c_{\rm NLO}(x, Q^2)$
%  and $\hat{R}^c_{\rm NNLO}(x, Q^2)$ are represented by solid, dashed and dotted gray curves, respectively.
  %Notation of curves is the same as in Fig.~1.
  %The experimental data are from ZEUS\cite{69} and H1\cite{71}.
%}
%\label{fig9}
%\end{center}
%\end{figure}

\end{document}